# A problem with the escort distribution representation of nonextensive statistical mechanics

Sumiyoshi Abe

*College of Science and Technology, Nihon University,*
*Funabashi, Chiba 274-8501, Japan*

It is pointed out that the Tsallis entropy functional represented in terms of the escort distribution is not concave if the entropic index $q$ is less than unity. It is emphasized that the escort distribution is a secondary object calculated from the basic original distribution.

In recent years, there has been growing interest in a nonextensive generalization of Boltzmann-Gibbs statistical mechanics. This formalism termed nonextensive statistical mechanics was initiated by Tsallis [1] by introducing a generalized entropy and was applied to various physical systems with long-range interactions, long-range memories, (multi)fractal structure and spatio-temporal complexity[1]. Among others, its recent impressive success in describing the phenomena of fully developed turbulence should be noticed [2,3].

Nonextensive statistical mechanics has a specific mathematical structure. It makes use of the concept of the escort distribution [4] to define expectation values of physical quantities. Since the escort distribution plays an essential role, some manipulations have been made to represent the theory completely in terms of it in the literature (which we do not specify here). The purpose of this short note is to call an attention to a problem regarding the concavity property of the entropy functional in the escort distribution representation. The escort distribution cannot be taken as a basic variable in maximum entropy calculus and is a secondary object calculated from the basic original distribution.

To make the point clear, it seems appropriate to first summarize the standard formalism of nonextensive statistical mechanics. This theory is based on the Tsallis

---

[1] A comprehensive list of references is currently available at

http://tsallis.cat.cbpf.br/biblio.htm

entropy

$$S_q[p] = \frac{1}{1-q}\left[\sum_{i=1}^{W}(p_i)^q - 1\right], \quad (1)$$

where $\{p_i\}_{i=1,2,\ldots,W}$ is the normalized probability distribution, $W$ is the number of accessible microscopic states of the system under consideration and $q$ is a positive parameter termed the index of nonextensivity. In the limit $q \to 1$, $S_q[p]$ converges to the ordinary Boltzmann-Shannon entropy $S[p] = -\sum_{i=1}^{W} p_i \ln p_i$. Difference of $q$ from unity indicates the degree of nonextensivity. An important point is that $S_q[p]$ is a concave functional for all positive values of $q$. To develop generalized canonical ensemble theory, we maximize the Tsallis entropy under the constraints on the normalization of the distribution and the expectation value of the system Hamiltonian $H$. Regarding the latter constraint, the authors of Ref. [5] introduced the concept of the normalized $q$-expectation value. Let $\varepsilon_i$ be the value of the Hamiltonian in the $i$th state of the system. Then, the generalized internal energy $U_q$ is given by

$$\sum_{i=1}^{W} P_i^{(q)} \varepsilon_i = U_q, \quad (2)$$

where $P_i^{(q)}$ is the escort distribution [4] defined as follows:

$$P_i^{(q)} = \frac{(p_i)^q}{\sum_{i=1}^{W}(p_i)^q}. \quad (3)$$

The functional to be maximized is given by

$$\Phi_q[p;\alpha,\beta] = S_q[p] - \alpha\left(\sum_{i=1}^{W} p_i - 1\right) - \beta\left(\sum_{i=1}^{W} P_i^{(q)} \varepsilon_i - U_q\right), \quad (4)$$

where $\alpha$ and $\beta$ are the Lagrange multipliers. Using variational principle, this functional is found to be maximized by the following distribution:

$$p_i = \frac{1}{\tilde{Z}_q(\beta)}\left[1 - (1-q)(\beta/c)(\varepsilon_i - U_q)\right]^{1/(1-q)}, \quad (5)$$

where

$$c \equiv \sum_{i=1}^{W} (p_i)^q, \quad (6)$$

$$\tilde{Z}_q(\beta) = \sum_{i=1}^{W} \left[1 - (1-q)(\beta/c)(\varepsilon_i - U_q)\right]^{1/(1-q)}. \quad (7)$$

From the normalization condition on $p_i$, follows the identical relation

$$c = \left[\tilde{Z}_q(\beta)\right]^{1-q}. \quad (8)$$

Using eqs. (5), (6) and (8), we find another expression for $\tilde{Z}_q(\beta)$:

$$\tilde{Z}_q(\beta) = \sum_{i=1}^{W} \left[1 - (1-q)(\beta/c)(\varepsilon_i - U_q)\right]^{q/(1-q)}. \quad (9)$$

In the limit $q \to 1$, the distribution in eq. (5) converges to the ordinary Boltzmann-Gibbs distribution $p_i = \exp(-\beta \varepsilon_i)/Z(\beta)$ with $Z(\beta) = \sum_{i=1}^{W} \exp(-\beta \varepsilon_i)$.

Now, let us consider the escort distribution representation. From eq. (3) and the normalization condition of $p_i$, we have

$$\sum_{i=1}^{W} (p_i)^q = \left[\sum_{i=1}^{W} \left(P_i^{(q)}\right)^{1/q}\right]^{-q}. \tag{10}$$

Therefore, the Tsallis entropy is rewritten as

$$\bar{S}_q[P^{(q)}] = \frac{1}{1-q}\left\{\left[\sum_{i=1}^{W}\left(P_i^{(q)}\right)^{1/q}\right]^{-q} - 1\right\}. \tag{11}$$

Now, in the escort distribtuion representation, the following functional is considered:

$$\Psi_q[P^{(q)}; \alpha, \beta] = \bar{S}_q[P^{(q)}] - \alpha\left(\sum_{i=1}^{W} P_i^{(q)} - 1\right) - \beta\left(\sum_{i=1}^{W} P_i^{(q)}\varepsilon_i - U_q\right). \tag{12}$$

The variation with respect to the escort distribution gives rise to

$$P_i^{(q)} = \frac{1}{\tilde{Z}_q(\beta)}\left[1 - (1-q)(\beta/c)(\varepsilon_i - U_q)\right]^{q/(1-q)}, \tag{13}$$

where $\tilde{Z}_q(\beta)$ is the same as that given in eq. (7) [or eq. (9)]. This is precisely equal to the escort distribution associated with $p_i$ in eq. (5). Therefore, apparently, the escort distribution representation yields the same result as the standard formalism if eqs. (3)

and (9) are used.

However, there are at least two points to be noticed here. First of all, the normalization condition on the escort distribution cannot be a constraint if the definition in eq. (3) is used. That is, $\sum_{i=1}^{W} P_i^{(q)} = 1$ becomes an identity in this case. Therefore, if $\sum_{i=1}^{W} P_i^{(q)} = 1$ is taken as a constraint, then the relation in eq. (3) has to be discarded and accordingly $P_i^{(q)}$ has to be regarded as a fundamental variable.

Secondly and more crucially, the Tsallis entropy in the representation in eq. (11) is seen to have a problem with the concavity condition, in general. In Fig. 1, we present plots of the second-order derivative, $f(r;q) \equiv \partial^2 \bar{S}_q[P^{(q)}] / \partial r^2$, with respect to $(r, q)$ in the case of the simplest two-state system: $P_1^{(q)} = r$, $P_2^{(q)} = 1 - r$ $(0 < r < 1)$. As clearly seen, $\bar{S}_q[P^{(q)}]$ is not concave, i.e. $f(r;q)$ changes its sign, if $0 < q < 1$, which is an important regime in nonextensive statistical mechanics. Thus the theory exhibits instability in this regime. In this context, it may also be worth pointing out the fact that nonextensive statistical mechanics with $0 < q < 1$ can give rise to the negative specific heat [6].

In conclusion, we have pointed out that that the escort distribution representation in nonextensive statistical mechanics has a problem regarding the concavity property of the entropy functional and the escort distribution cannot be taken as a basic variable in maximum entropy calculus. The escort distribution in nonextensive statistical mechanics is a secondary object, which should be constructed from the basic original distribution $\{p_i\}_{i=1, 2, \text{L}, W}$.

**Acknowledgment**

This work was supported in part by the GAKUJUTSU-SHO Program of College of Science and Technology, Nihon University.

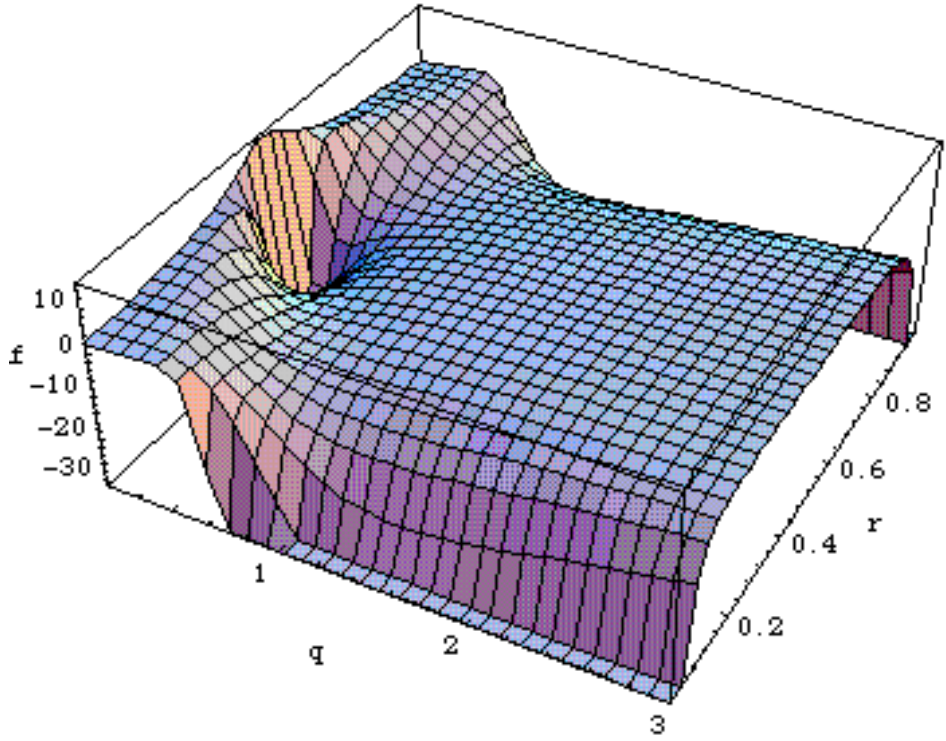

Fig.1 Plots of $f(r;q) \equiv \partial^2 \overline{S}_q[P^{(q)}] / \partial r^2$ for the two-state system,

$P_1^{(q)} = r$, $P_2^{(q)} = 1 - r$ $(0 < r < 1)$, with respect to $(r, q)$.